\documentclass[aps,prl,showpacs,twocolumn,floats]{revtex4-1}
\usepackage{amssymb}
\usepackage{amsbsy}
\usepackage{amsmath}
\usepackage{epsfig}
\usepackage{amsfonts}
\usepackage{graphicx}
\usepackage{amssymb}
\usepackage{upgreek}
\usepackage{dsfont}
\usepackage{bm}
\usepackage{color}
\usepackage{hyperref}

\begin{document}

\title {Photoinduced Chern insulating states in semi-Dirac materials}
\author{Kush Saha}
\affiliation{Department of Physics and Astronomy, University of California, Irvine, California 92697, USA\\
 California Institute for Quantum Emulation, Santa Barbara, California 93106, USA }
\date{\today}
\begin{abstract}
Two-dimensional (2D)  {\it semi-}Dirac materials are characterized by a quadratic dispersion in one direction and a linear dispersion along the orthogonal direction. We study the topological phase transition in such 2D systems in the presence of an electromagnetic field. We show that a Chern insulating state emerges in a {\it semi}-Dirac system with two {\it gapless} Dirac nodes in the presence of light. In particular, we show that the intensity of a circularly polarized light can be used as a knob to generate topological states with nonzero Chern number. In addition, for fixed intensity and frequency of the light, a {\it semi}-Dirac system with two {\it gapped} Dirac nodes with trivial band topology can reveal the topological transition as a function of polarization of the light. 
\end{abstract}
\maketitle

{\em Introduction-} 
Recent years have witnessed unprecedented theoretical and experimental advances in the field of materials whose  
low-energy excitations are described by Dirac fermions \cite{novo, castro,fu,roy,hasan0, books}. 
Such materials with a bulk {\it gapped} spectrum (e.g., 2D HgTe/CdTe quantum wells, 3D Bi$_2$Se$_3$) 
may exhibit exotic electronic properties such as a quantum spin Hall effect\cite{bernevig}, topological magneto-caloric effect\cite{qi}, etc.
Likewise, {\it gapless} or semi-metallic Dirac materials (e.g., 2D graphene, 3D Weyl metal) may reveal many unconventional properties such as a   
quantum Hall effect at room temperature (particularly in graphene)\cite{novo1}, chiral anomaly induced negative magnetoresistance\cite{huang} (particularly in a Weyl metal), and many more\cite{castro,xu}. Due to their exotic properties, on the one hand, there has been a considerable effort to search for new materials with Dirac-like properties. This has led to several proposals to realize a new class of materials with specific 2D Dirac dispersion: parabolic in one direction and linear in the perpendicular direction. The materials or systems that can host such {\it semi}-Dirac (SD) dispersion are TiO$_2$/V$_2$O$_3$\cite{padro}, BEDT-TTF$_2$I$_3$ salt under pressure\cite{kata}, hexagonal lattices in the presence of a magnetic field\cite{diet}, and photonic system\cite{yu}. Due to their unusual dispersion, they may exhibit exotic properties\cite{del,banerjee} in contrast to graphene or conventional 2D systems.

On the other hand, there is a race to engineer new materials with topological properties due to the lack of 
natural materials with such properties. The static controllable parameters
by which one can induce topological phases in intrinsically nontopological materials 
include pressure\cite{bahmy}, doping\cite{hasan}, disorder\cite{li}, temperature\cite{ion}, and a few more\cite{diss}. However, most of these tools lack
a continuous and high degree of controllability. Recently, it has been shown and later verified by experiment that time-dependent perturbations can induce topological phases in an intrinsically nontopological insulating material \cite{taki,linder, leon, chan, yavir, gedik,  tanaka, kitagawa, peter, adolfo,calvo,molina,wang, narayan1} . Also, it has been shown that polarized light can open a gap in semi-metallic systems, allowing them to host topologically protected gapless modes responsible for robust transport properties. Indeed, {\it circularly} polarized light opens a gap in a gapless Dirac system such as graphene, and may give rise to a photoinduced Hall current without applying any magnetic field\cite{taki}.  Thus semi-metallic graphene becomes a Chern insulator (CI) in the presence of electromagnetic radiation, making it potential candidate for spintronics or transistors. 

\begin{figure}
\includegraphics[width=0.99\linewidth]{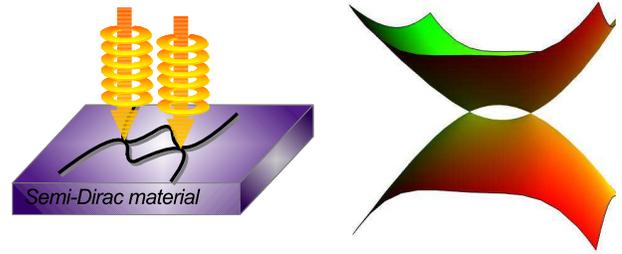}
\caption{a) Schematics of shining polarized light on the surface of a SD material. The size of the sample is considered to be smaller than the wavelength of the light. b) Energy spectrum of the SD Hamiltonian in Eq.~(\ref{ham0}) with two nodal points at $(\pm \sqrt{\delta_0/\alpha},0)$. Note that, this dispersion is gapped if $\delta_0<0$. }
\label{fig1}
\end{figure}

This raises an interesting question if {\it semi}-Dirac systems can carry a nonzero Chern number in the presence of polarized light. In this article, we have answered this affirmatively. For {\it circularly} polarized light,  we show light can open a gap and induce topological states in a semi-metallic SD system. In particular, the strength of the polarized light can be used as a knob to drive a topological transition. In addition, we show that a SD system with a trivial insulating phase can mimic the Haldane model as a function of polarization of the light. Finally, we investigate photoinduced anomalous Hall conductivity for different strengths of the incident beam. Note that a similar system was studied \cite{narayan}  recently in the presence of laser light but overlooked the interesting topological transition as presented here.

{\em Model-}
The minimal model Hamiltonian describing low-energy electronic bands of a two-dimensional {\it semi}-Dirac material is\cite{del,banerjee}
\begin{align}
H_0(\bf k)={\bf d(\bf k)}\cdot {\bf \sigma},
\label{ham0}
\end{align}
where ${\bf \sigma}=(\sigma_x,\sigma_y,\sigma_z)$ are the Pauli matrices in  pseudospin space, $ {\bf d (\bf k)}=(\alpha k_x^2-\delta_0,v k_y,0)$,
where ${\bf k}=(k_x, k_y)$ is the crystal momentum,  $\alpha$ is the inverse of quasiparticle mass along $x$, $v$ is the Dirac velocity along $y$, and $\delta_0$ is the gap parameter. The energy eigenvalues are given by
\begin{align}
E^{\pm}_{k_x,k_y}=\pm \sqrt{(\alpha k_x^2-\delta_0)^2+v^2 k_y^2},
\label{eval}
\end{align}
where $\pm$ denote the conduction and valence band, respectively. For $\delta_0=0$, the spectrum is gapless with linear dispersion along $k_y$. For $\delta_0<0$, it is a gapped system with trivial insulating phase. For $\delta_0>0$, we obtain two gapless Dirac points at $(\pm \sqrt{\delta_0/\alpha},0)$. The corresponding energy spectrum is shown in Fig.~(\ref{fig1})b. Note that, this model obeys effective time-reversal ($\Theta=\mathcal {K}$), particle-hole ($\mathcal{P}=\sigma_y$) and chiral symmetries ($\mathcal{C}=\sigma_y \mathcal {K}$), where $\mathcal{K}$ is the complex conjugation operator. It is worth mentioning that, in typical SD systems, the gapless modes are protected  by mirror symmetry along some symmetry line\cite{huang1} and cannot be destroyed without breaking that symmetry.

\begin{figure}
\includegraphics[width=0.99\linewidth]{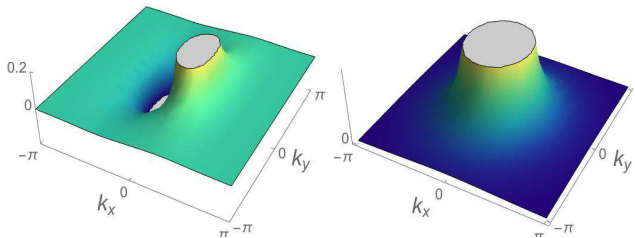}
\caption{Berry curvature of a gapped {\it semi}-Dirac Hamiltonian in the presence of perturbation $\delta H=m_z\sigma_z$. The integral of the Berry curvature over the BZ turns out to be zero since $\Omega(-k_x,-k_y)=-\Omega(k_x,k_y)$ as a manifestation of time-reversal symmetry.  b) Same plot as (a) for 2D linear Dirac dispersion \cite{dirac}. In this case, the perturbation breaks time-reversal symmetry and leads to a nonzero Chern number\cite{dirac}, as is evident from the Berry curvature.}
\label{fig2}
\end{figure}

{\em Berry curvature and Chern number-}
Eq.~(\ref{ham0}) represents a generic Hamiltonian of a two-level system. Its simple form in terms of ``vector field" ${\bf d(k)}$ allows us to
write the Berry curvature as 
\begin{align}
{\bf \Omega}({\bf k})=\frac{1}{2}\frac{\bf d(k)}{|\bf d( k)|^3}.
\end{align}
Since the Chern number is defined as an integral of Berry curvature over the 2D Brillouin zone (BZ), it can be expressed as 
\begin{align}
C=\int d{\bf k}\Omega({\bf k})={1\over 2\pi}\int  d{\bf k}~ {1\over 2 |\bf d|^3} {\bf d} .(\partial_{k_x}{\bf d}\times \partial_{k_y}{\bf d}),
\end{align}
where $\partial_{x}$ is the partial derivative with respect to $x$.
Since the $z$-component of ${\bf d}$ in Eq.~(\ref{ham0}) is zero, the Berry curvature is zero everywhere except at the gapless Dirac points where it diverges. The perturbation $\delta H=m_z \sigma_z$ opens a gap and leads to a nonzero local Berry curvature,
\begin{align}
\Omega(k_x,k_y)=\pm \frac{2\alpha v m_z k_x}{(E_{k_x,k_y})^3}.
\label{berry}
\end{align} 
Note that $\Omega(k_x,k_y)$ is asymmetric in $k_x$ for constant $m_z$. Thus, the Chern number $C$ is obtained to be zero, as also apparent from the Berry curvature shown in Fig.~(\ref{fig2})a. The above perturbation therefore cannot change the band topology in the SD system. This is in contrast to {\it half} semimetals  with 2D linear Dirac bands near the Fermi level, where a mass gap induced by spin-orbit coupling leads to a nonzero Chern number\cite{dirac,poly}. It is worth pointing out that, in graphene, the perturbation $m_z\sigma_z$, namely the ``Semenoff mass" leads to a trivial insulating phase, while perturbation $m_z\sigma_z\tau_z $, namely the ``Haldane mass" gives rise to non-trivial band topology, where $\tau_z$ denotes valley the degrees of freedom\cite{haldane}.  

{\em Floquet theory-}
We now investigate the effect of time-dependent radiation generated by a radio frequency source or a laser on the semi-metallic phase of 
the SD Hamiltonian in Eq.~(\ref{ham0}). The light field ${\bf A}(t)=A_0(\sin(\omega t),\sin(\omega t+\phi) )$ minimally couples to the system via the momentum ${\bf k}\rightarrow {\bf k}+e{\bf A}(t)$, where $\phi$ is the polarization of the light, $e$ is the electric charge, $\omega$ is the frequency of the light, and $A_0$ is the strength of the applied field. Note that we neglect spatial dependence of the light field considering the fact that the wavelength of the light field is large compared to the sample size.

Then, in the presence of a light field, Eq.~(\ref{ham0}) can be read off as
\begin{align}
H({\bf k},t)=H_0({\bf k})+H_{1}({\bf k},t)\sigma_x+H_{2}({\bf k},t)\sigma_y,
\end{align}
where $H_{1}({\bf k},t)=\frac{\alpha e^2 A_0^2}{2}(1-\cos(2\omega t))+2\alpha e A_0 k_x \sin(\omega t)$,
and $H_{2}({\bf k},t)=ve A_0 \sin(\omega t+\phi)$. 

Under this time-dependent Hamiltonian, the quantum state evolves as $\Psi(t)=U(t,t_0)\Psi(t_0)$, where 
$U(t,t_0)=\tau \exp(-\frac{i}{\hbar}\int_{t_0}^{t} dt' H({\bf k},t')) $, where $\tau$ is the time-ordering operator, and $t_0$ is the initial time of the perturbation.  Using the  Floquet formalism, we define an effective static Hamiltonian, namely Floquet Hamiltonian after a full time period $T$  as 
\begin{align}
H_F ({\bf k})=\frac{i}{\hbar T} \ln(U(T)),
\end{align}  
where we have assumed $t_0=0$. Following Ref.~[\onlinecite{anatoli}], we obtain the Floquet Hamiltonian up to first order in inverse frequency,
\begin{align}
H_F({\bf k} )\simeq H_F^{0}+H_{F}^1,
\label{floq}
\end{align}
where 
$H_{F}^0$ is obtained by time-averaging of $H({\bf k},t)$:
\begin{align}
H_{F}^0=\frac{1}{T}\int_{0}^T H({\bf k},t )=H_0({\bf k} )+\frac{\alpha e^2 A_0^2 }{2}\sigma_x,
\end{align} 
and 
\begin{eqnarray}
H_{F}^1&=&\frac{1}{2T i \hbar}\int_{0}^{T}dt_1\int_{0}^{t_1}[H({\bf k},t_1),H({\bf k},t_2)] dt_2\nonumber\\
&=&(m_0+\beta k_x^2+\gamma k_x+\eta k_x k_y)\sigma_z,
\end{eqnarray} 
where $m_0=\frac{eA_0 v}{\hbar \omega}(\alpha e^2 A_0^2-2\delta_0)\cos(\phi)$, $\beta=\frac{eA_0 v}{\hbar \omega}2\alpha \cos(\phi)$,
$\gamma=\frac{e^2A_0^2 v}{\hbar \omega}2\alpha\sin(\phi)$ and $\eta=-\frac{4\alpha eA_0 v}{\hbar\omega}$. 
Note that this is a high-frequency expansion when only one photon is considered for the process. 
We recognize $H_{F}^1$ as a gap-opening mass term in Eq.~(\ref{ham0}). As already discussed, $H_0({\bf k})$ represents the Hamiltonian of a trivial insulator in the presence of a gap induced by a constant mass term. However, we will see that the momentum dependent term in $H_F^1$ plays a crucial role in revealing the band topology in these systems. 

Combining $H_F^{0}$ and $H_{F}^1$, Eq.~(\ref{floq}) yields
\begin{align}
H_F({\bf k} )={\bf n({\bf k})}.{\bf \sigma},
\end{align}
with 
\begin{align}
{\bf n({\bf k})}=(\alpha k_x^2+\delta_1,v k_y, m_{\rm eff} ),
\label{newd}
\end{align}
where $\delta_1=\frac{\alpha e^2 A_0^2}{2}-\delta_0$, $m_{\rm eff}=m_0+\beta k_x^2+\gamma k_x+\eta k_x k_y$. Note that the incident beam shifts the position of the Dirac nodes. Notice also, that $m_{\rm eff}$ breaks the effective time-reversal symmetry as $\Theta m_{\rm eff}(k_x,k_y)\Theta^{-1}\neq m_{\rm eff}(-k_x,-k_y)$.

With the new vector field ${\bf n(k)}$, we define the Chern number for the Floquet Hamiltonian as
 \begin{align}
 C_{\rm F}=&{1\over 2\pi}\int  d{\bf k}~ {1\over  2|\bf n(k)|^3} {\bf n (k)} .(\partial_{k_x}{\bf n(k)}\times \partial_{k_y}{\bf n(k)}),
\label{eberi}
 \end{align}
where $|n({\bf k})|=\pm\sqrt{(\alpha k_x^2+\delta_1)^2+v^2 k_y^2+m_{\rm eff}^2}$. 

\begin{figure}
\includegraphics[width=0.99\linewidth]{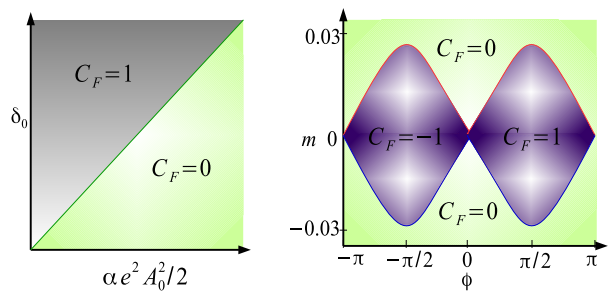}
\caption{a) Schematic phase diagram of the photoinduced Chern number ($C_F$) in the $\delta_0-\alpha e^2 A_0^2/2$ plane. The diagonal lines correspond to $\delta_0=\alpha e^2A_0^2/2$, where $C_F$ is ill-defined as the spectrum is gapless. b) Phase diagram of $C_F$ in the $m-\phi$ plane for a gapped {\it semi}-Dirac system as discussed in the text. The red ($m_{c_1}>0$) and blue lines ($m_{c_2}<0$) correspond to a gapless spectrum due to the gap closing at only one Dirac point.}
\label{fig4}
\end{figure} 

{\em Chern insulating states in {\rm gapless} SD systems-} Together with Eq.~(\ref{newd}) and the integral part of Eq.~(\ref{eberi}), we obtain
\begin{align}
\Omega(k_x,k_y)= \frac{v}{2|n(\bf k)|^3}\left[\gamma(\alpha k_{x}^2-\delta_1)-\eta k_y(\alpha k_x^2+\delta_1)\right].
\label{case}
\end{align} 
It is apparent that $\Omega(k_x,k_y) \neq - \Omega(-k_x,-k_y)$ as a consequence of broken time-reversal symmetry.  
We would like to point out that the numerator of $\Omega(k_x,k_y)$ does not contain any term involving $m_0$. However, the presence 
of such a term in the numerator can have important consequences in revealing the band topology as will be evident below.  

For {\it circularly} polarized  ($\phi=\pm \pi/2$) light, and for $\delta_0>\alpha e^2A_0^2/2$,  the gapless Dirac nodes become gapped with their new position at $(\pm \sqrt{\frac{|\delta_1|}{\alpha}-\frac{\gamma^2}{2\alpha^2}},0)$.
Then the total Berry flux for this gapped spectrum is found to be $\pm2\pi {\rm sgn}(\gamma)$, where $\pm$ correspond to the positive and negative energy spectrum, respectively. Note that  {\it left} ($\phi=-\pi/2$) or {\it right} circularly ($\phi=+\pi/2$) polarized light determines the sign of $\gamma$. The $2\pi$ flux comes from the fact that the energy spectrum has two Dirac-like dispersions, and each of those contributes $\pm\pi$ to the total flux in contrast to $2\pi$ flux in bilayer graphene due to quadratic band touching. Consequently, two gapped Dirac nodes make a contribution of $\pm(1/2+1/2)=\pm 1$ to $C_F$. The total $C_F$ is obtained by summing over the contribution coming from all Dirac nodes in the BZ. For typical semi-Dirac materials, there are four symmetry-related momenta \cite{diet}, which, in turn, give rise to a total of 8 Dirac nodes, hence a total $C_F=\pm 4$. However, for simplicity, we restrict ourselves to a system with only two Dirac nodes within the BZ\cite{cand}.  
  
For $\delta_0=\alpha e^2A_0^2/2$, i.e., $\delta_1=0$, the spectrum is gapless. For $\delta_0<\alpha e^2A_0^2/2$, the spectrum becomes gapped again. In this case, $C_F$ turns out to be zero. Thus varying intensity of the incident light, we can drive topological transition in a gapless SD system (cf. Fig.~\ref{fig4}a). Note that, unlike the case of gapped trivial insulating phase induced by constant mass term (as discussed earlier), the Berry curvature $\Omega(k_x,k_y)$ is not asymmetric either in $k_x$ or $k_y$. We therefore expect to have finite anomalous Hall conductivity at finite doping ($\epsilon_F\ne 0$). This will be discussed shortly.

{\em Chern insulating states in {\rm gapped} SD systems-} 
The objective of this section is to explore topological phenomena in a SD system with two {\it gapped} Dirac nodes within the BZ. The low-energy model Hamiltonian is given by
\begin{align}
H_{\rm g}({\bf k})=H_0({\bf k})+m\sigma_z.
\end{align}
As mentioned earlier, this Hamiltonian represents a trivial insulator with zero Chern number as a consequence of an asymmetric Berry curvature. The presence of polarized light leads to $n_{\rm g}({\bf k})=(\alpha k_x^2+\delta_1-m\xi,v k_y+m\chi k_x, m+m_{\rm eff} )$, where $\xi=2eA_0  v\cos(\phi)/\hbar\omega$ and $\chi=4eA_0 \alpha/\hbar\omega$. Note that the position of the Dirac nodes moves along both $k_x$ and $k_y$ directions as opposed to the case discussed in the preceding section. Note also, that the movement depends on the polarization of the light. 

With this  $n_{\rm g}({\bf k})$, the numerator of the Berry curvature turns out to be dependent on $m$, and also on $\cos(\phi)$\cite{sup}, in contrast to the previous case where the term involving $m_0$ vanishes identically in the numerator.  Consequently, for fixed intensity and frequency of the light, we can mimic the phase diagram of the Haldane model by varying  $m$ with the polarization of the light. The corresponding plot is shown in Fig.~(\ref{fig4})b.  Along the curve lines ( $m_{c_1} \simeq m_{c_2}=m_c$), one Dirac node closes while the other remains open, hence the spectrum becomes gapless. Note that $m_c$ depends on the intensity, frequency, and  polarization of the light. For {\it circularly} polarized light, and for $\delta_1<0$, we obtain $m_c(\phi=\pi/2)=\pm2e^2A_0^2 v\hbar\omega \sqrt{\alpha|\delta_1|}/(16 e^2A_0^2\alpha\delta_1-\hbar^2\omega^2)$. For any other polarization, $m_c$ can be obtained numerically for fixed $A_0$ and $\omega$. When $|m|<|m_c|$, the mass gaps at two Dirac nodes are opposite in sign, hence $C_F=\pm {\rm sgn}(\phi)$. In contrast, for $|m|>|m_c|$, the mass gaps at two Dirac nodes turn out to be the same in sign. Thus we obtain $C_F=0$, indicating a transition from a topological to a trivial phase as a function of polarization of the light.
 
Before ending this section, we would like to point out that a gapped SD system for $\delta_0<0$ in Eq.~(\ref{ham0}) does not host any interesting topological states in the presence of light. Thus the presence of gapless or gapped Dirac nodes (when $\delta_0>0$) in the field-free Hamiltonian of SD materials plays a crucial role in revealing the topological transition in the presence of light. In sum, this  reflects a non-trivial interplay between light and band spectrum. Finally, we comment on the lattice model which is crucial to compute the Chern number correctly. In doing so, we consider the lattice Hamiltonian, $H=(1-2\cos(k_x))\sigma_x+k_y\cos(k_x)\sigma_y$, which has two Dirac nodes within the 2D BZ. 
Note that, this model Hamiltonian can be obtained from hexagonal lattice geometry with appropriate band parameters\cite{montabaux}. Then 
the perturbation $\delta H=\gamma\sin(k_x)\sigma_z$ reveals the same topological properties as discussed in the preceding sections.

\begin{figure}
\includegraphics[width=0.99\linewidth]{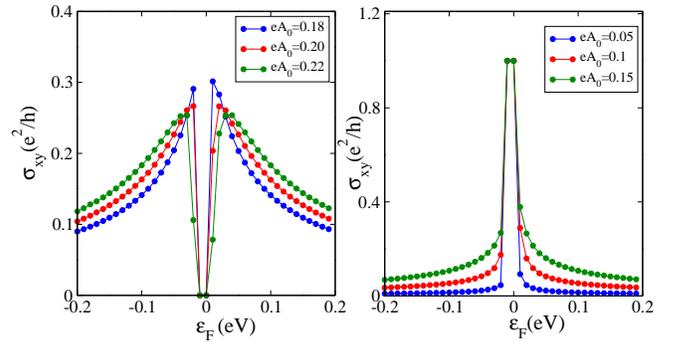}
\caption{a) Anomalous Hall conductivity $\sigma_{xy}$ for a trivial insulating phase with nonzero Berry curvature as a function of $\epsilon_F$. It is evident that $\sigma_{xy}$ increases with the intensity $A_0$. However, as soon as $\epsilon_F$ reaches the gap, it drops to zero as a consequence of zero Chern number for $\delta_1>0$. b) For $\delta_1<0$, we see the conductivity is quantized in units of $e^2/h$ if $\epsilon_F$ lies within the gap. Away from the gap, we obtain a very small contribution to conductivity since $\gamma$ is small for weak intensity $A_0$.}
\label{fig5}
\end{figure}

{\em Anomalous Hall conductivity-} 
Unlike quantum Hall conductivity, the anomalous Hall conductivity does not require any external magnetic field or nonzero Chern number. The local Berry curvature can contribute to conductivity even at zero temperature for finite Fermi energy, $\epsilon_F$. The Hall conductivity is defined as
\begin{align}
\sigma_{xy}^{\pm}=\pm\frac{e^2}{\hbar}\int \frac{d{\bf k}}{(2\pi)^2} f(E^{\pm}_{k_x,k_y})\Omega(k_x,k_y),
\label{condu}
\end{align} 
where $f(\epsilon)$ is the Fermi-Dirac (FD) distribution function, $f(\epsilon)=(1+e^{(\epsilon-\epsilon_F)/K_B T})^{-1}$, where $T$ is the temperature. The total conductivity is the sum of two $\sigma_{xy}^{\pm}$s. Note that Eq.~(\ref{condu}) is best valid for equilibrium dc Hall conductivity. In nonequilibrium,  the expression for $\sigma_{xy}$ differs due to the nonuniversal nature of the FD distribution, which, in turn, depends on the details of the systems, its environment, etc \cite{taki,luis,mitra}. However, we approximate $\sigma_{xy}$ by its equilibrium value based on the fact that the energy of the perturbing field exceeds any other energy scales of the system\cite{mitra}.  At zero temperature,  $f(E^{\pm}_{k_x,k_y})=\Theta(E^{\pm}_{k_x,k_y}-\epsilon_F)$. For a typical semi-Dirac material, we take $v=0.65 eV \r{A}$, $\alpha=0.75 eV \r{A}^2$, $\delta_0=0.01 eV$. The typical phonon energy that is used in experiment is roughly $0.25 eV$  with $eA_0=0.01-0.2 \r{A}^{-1}$. Using these parameters, we compute the anomalous conductivity, particularly for a {\it gapless} SD system when $\delta_0<\alpha e^2A_0^2/2$. Fig.~(\ref{fig5})a shows the dependence of $\sigma_{xy}$ on $\epsilon_F$ and $A_0$. Note that $\epsilon_F=0$ corresponds to Fermi energy in the middle of the gap. As $\epsilon_F$ approaches to the gap, the conductivity gradually increases, and drops to zero as $\epsilon_F$ reaches the gap. Notice that, a measurable conductivity contribution can be obtained if $\epsilon_F$ lies close to the gap. Also, the magnitude of Hall conductivity varies with the strength of the vector potential provided that $v e A_0<<\hbar\omega$ as a valid criterion for Floquet expansion.
Fig.~(\ref{fig5})b shows zero temperature conductivity for $\delta_0>\alpha e^2A_0^2/2$. As expected, the conductivity is quantized if $\epsilon_F$ lies within the gap. However, away from the gap the conductivity appears to be small. This is because of the low intensity set by the condition  $\delta_0>\alpha e^2A_0^2/2$. 

The predicted photoinduced chiral nature of the 2D {\it semi-}Dirac materials can be verified by dynamical Hall measurement using a crossed ac electric and magnetic field of circularly polarized light\cite{korch}. In addition, since the intensity of the electromagnetic radiation is easily controllable in experiment, the signature of transition from a topological to a trivial phase can be tested in this transport experiment, provided that $\delta_0$ in the band spectrum is positive.

{\em Conclusion-} We have shown that light can induce topological states in an intrinsically trivial 
{\it semi}-Dirac materials. For circularly polarized light, the intensity of the light can be used to induce topological phases in a semi-metallic {\it semi}-Dirac system.
On the other hand, for fixed intensity and frequency, the polarization of light can drive a topological transition in a gapped {\it semi}-Dirac system.   
In addition, the intensity of light can be used to tune the anomalous Hall conductivity to measurable values for Fermi energy lying either in the conduction or valence bands. 

KS thanks M. Kolodrubetz, S. A. Parameswaran and I. Garate for useful discussions. This work is funded by NSF Grant DMR-1455366 and the California Institute for Quantum Emulation, supported by a President's Research Catalyst Award (CA-15-327861) from the University of California Office of the President.


\begin{thebibliography}{99}
\bibitem{cand} 2D organic metals or half-metallic systems may be promising candidates that can host only two Dirac nodes near the Fermi level.
\end{thebibliography}
\end{document}